\input{aipcheck}

\pdfoutput=1
\documentclass[
    ,final            % use final for the camera ready runs
  ]
  {aipproc}

\layoutstyle{6x9}

\def\beq{\begin{equation}}
\def\eeq{\end{equation}}
\def\beqn{\begin{eqnarray}}
\def\eeqn{\end{eqnarray}}

\newcommand{\ntwo}{${\mathcal N}=2\,$}

\newcommand{\pt}{\partial}

\newcommand{\gsim}{\lower.7ex\hbox{$
\;\stackrel{\textstyle>}{\sim}\;$}}
\newcommand{\lsim}{\lower.7ex\hbox{$
\;\stackrel{\textstyle<}{\sim}\;$}}

\newcommand{\p}{\partial}

\def\slashed#1{\setbox0=\hbox{$#1$}             % set a box for #1
   \dimen0=\wd0                                 % and get its size
   \setbox1=\hbox{/} \dimen1=\wd1               % get size of /
   \ifdim\dimen0>\dimen1                        % #1 is bigger
      \rlap{\hbox to \dimen0{\hfil/\hfil}}      % so center / in box
      #1                                        % and print #1
   \else                                        % / is bigger
      \rlap{\hbox to \dimen1{\hfil$#1$\hfil}}   % so center #1
      /                                         % and print /
   \fi}                                        %

\newcommand{\D}{{\mathcal D}}

\begin{document}

\begin{flushright}
FTPI-MINN-09/32, UMN-TH-2812/09
\end{flushright}

\title{Non-Abelian Strings and Axions\footnote{Talk at the Workshop {\sl Axions 2010}, in honor
of the 60$^{th}$ birthday of Pierre Sikivie, University of Florida, January 15--17, 2010.}}


\author{M. Shifman}{
  address={William I. Fine Theoretical Physics Institute, University of Minnesota,
  Minneapolis, MN 55455}
}

\begin{abstract}
Axion-like fields can have a strong impact on non-Abelian strings. 
I discuss axion connection to such strings and its implications in two cases: (i) axion 
localized on the strings, and (ii) axions propagating in the four-dimensional bulk.
\end{abstract}

\maketitle

\section{Introduction}
\label{intro}

\begin{figure}
  \includegraphics[height=.45\textheight]{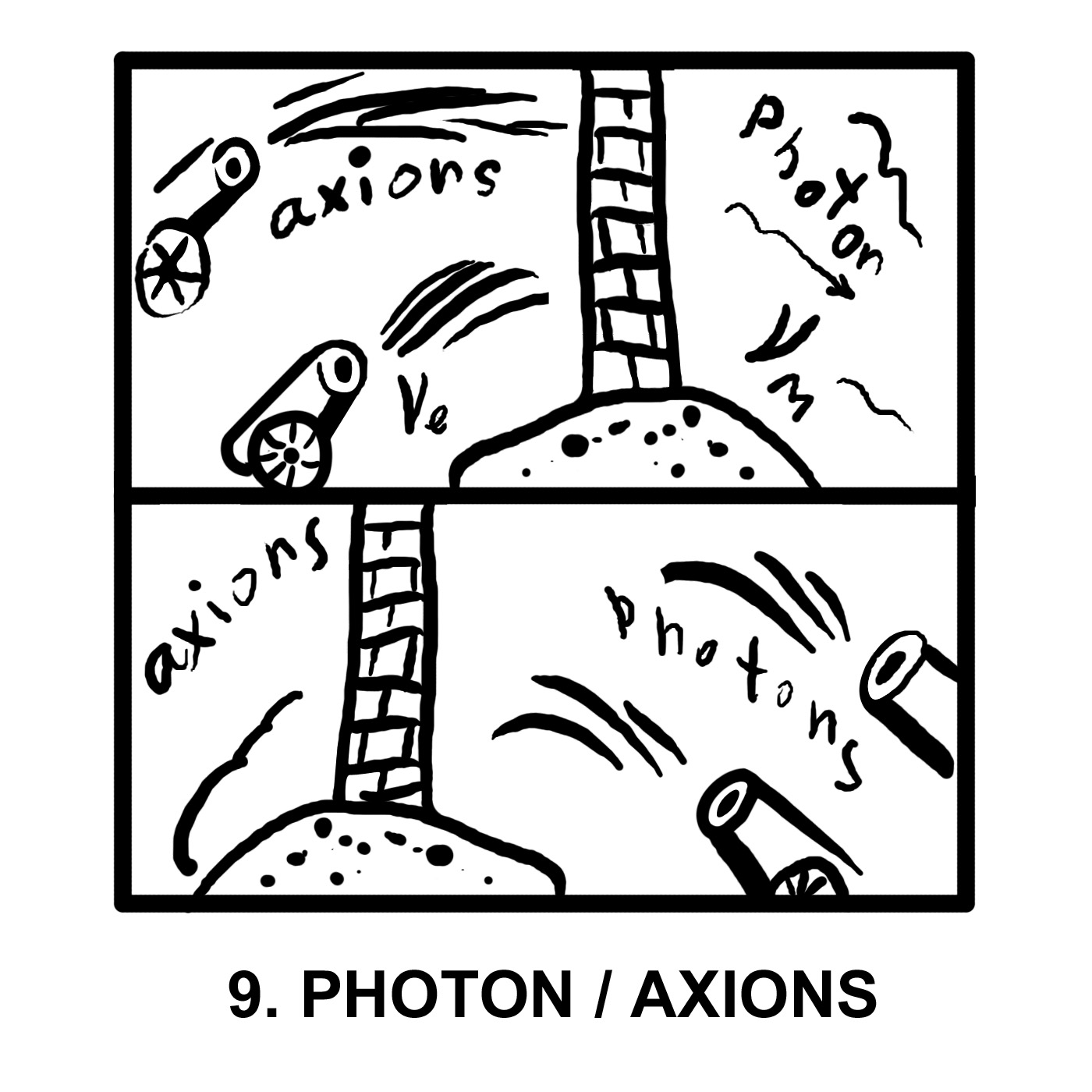}
  \caption{Axion research on industrial basis.}
  \label{loto}
\end{figure}

\begin{figure}
  \includegraphics[height=.27\textheight]{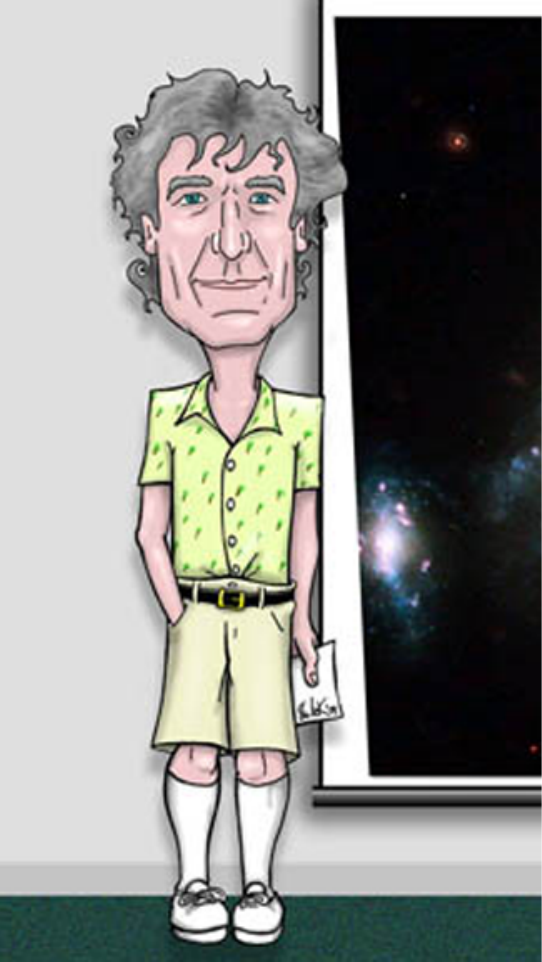}
  \caption{Pierre Sikivie, the Axionman.}
  \label{PA}
\end{figure}

Axions were introduced by Weinberg \cite{W1} and Wilczek \cite{W2}
in a bid to save naturalness of $P$ and $T$ parity conservation in QCD.
Shortly after, the axion construction evolved into  ``an invisible axion''
of the first \cite{K,svz} or the second \cite{Zh,ZF} kind. Moreover,
already in the early days of string theory people realized that axion-like particles
are  an unavoidable feature of string theory and are abundant. Since then, axions acquired a life of their own
(Fig.~\ref{loto}),
a big part of which is due to Pierre's unconditional commitment to this 
fundamental topic which goes as far as participation 
in experimental axion searches starting from the early stages of experiment design! 
I think it is fair to say that 
he shaped research in this area.
Pierre is the true and ultimate Axionman (Fig.\ref{PA}). Happy birthday, Pierre,
and many new findings in your exciting career!

In the last 10 years  I was only marginally connected with
 the development of ideas in the area of the axion phenomenology. The last active 
 effort in this direction was a review paper \cite{GabS}, written with Gregory Gabadadze,
 in which we explored consequences of  a newly acquired
knowledge of nonperturbative aspects of the QCD vacuum in axion physics.
Needless to say, such great occasion as Pierre's birthday calls for presentation of
fresh results. My current work is focused on non-Abelian strings, a construction 
which emerged recently~\cite{HT1,ABEKY,SYmon,HT2} (for a detailed review see  \cite{Trev}).
These strings could play the role of the cosmic strings \cite{Hashimoto},
which would be a very appropriate topic today,
but, unfortunately, this idea is not yet fully implemented
in viable phenomenological constructions. 
Therefore, 
I will talk today about a kind of axions which serve as a theoretical laboratory
in the explorations of flux tubes (strings) and other topological solitons, rather
than the ``real'' axions which 
are likely to be a part of our world (the favorite object of Pierre).
Most of the results to be reported today were obtained with Gorsky and Yung  \cite{Gorsky}.

\section{Briefly on non-Abelian strings}
\label{nas}

\begin{figure}
  \includegraphics[height=.22\textheight]{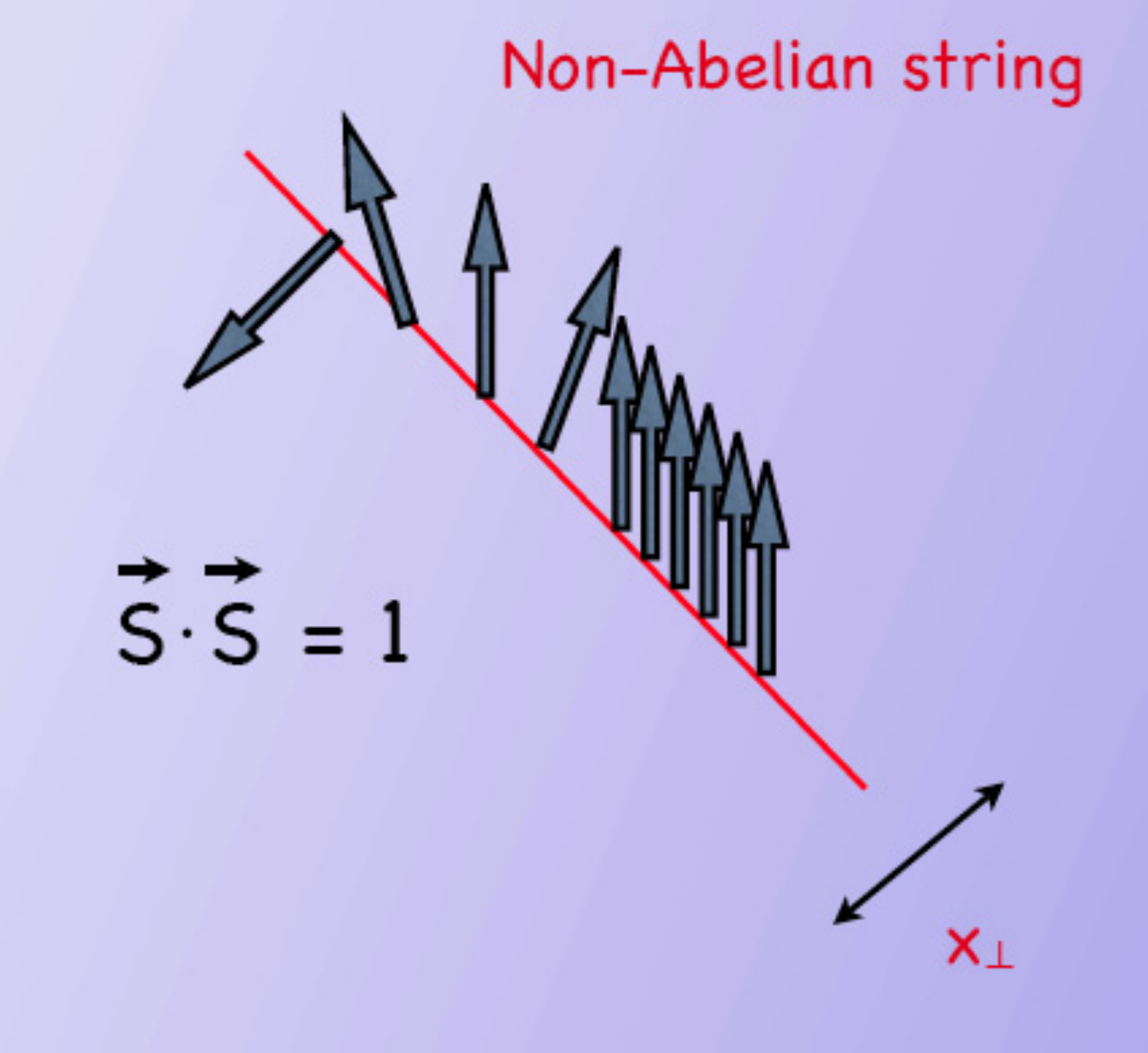}
  \caption{Orientational moduli on the string world sheet are depicted by arrows.}
  \label{f2}
\end{figure}

The Abrikosov flux tube (string) in the Abelian U(1) gauge theory is known\,\footnote{Relativistic generalization
was given in \cite{ANO}.}
from the 1950's \cite{Abri}. What is the difference between our good old acquaintance,
the Abrikosov string, and the new arrival, non-Abelian string? Of course, the non-Abelian
strings are usually found in non-Abelian gauge theories, but this is not their main 
defining feature. Of most importance is the fact that additional moduli -- (classically massless) fields
describing internal degrees of freedom -- exist on the string world sheet.
The most popular example refers to orientational moduli localized on the string \cite{HT1,ABEKY,SYmon,HT2}.
If the bulk theory has the U$(N)$ gauge symmetry and SU$(N)$ flavor symmetry, 
with the appropriate choice of the Higgs sector
\cite{Gorsky2}, the bulk theory is fully Higgsed, while still preserving a color-flavor locked 
global SU$(N)$ symmetry. The latter is broken down to
SU$(N-1)\times {\rm U}(1)$ on any given string solution. As a result, moduli living in a coset 
space ${\rm SU}(N)/({\rm SU}(N-1)\times {\rm U}(1))$ emerge.
Their interaction   is described by two-dimensional
CP($N-1)$ model (Fig.~\ref{f2}).

\vspace{2cm}

The simplest example is provided by the following (nonsupersymmetric) model:
\begin{eqnarray}
S &=& \int {\rm d}^4x\left\{\frac1{4g_2^2}
\left(F^{a}_{\mu\nu}\right)^{2}
+ \frac1{4g_1^2}\left(F_{\mu\nu}\right)^{2}
 \right.
 \nonumber\\[3mm]
&+&
 {\rm Tr}\, (\nabla_\mu \Phi)^\dagger \,(\nabla^\mu \Phi )
+\frac{g^2_2}{2}\left[{\rm Tr}\,
\left(\Phi^\dagger T^a \Phi\right)\right]^2
 +
 \frac{g^2_1}{8}\left[ {\rm Tr}\,
\left( \Phi^\dagger \Phi \right)- 2\xi \right]^2 
 \nonumber\\[3mm]
 &+&\left.
 \frac{i\,\theta}{32\,\pi^2} \, F_{\mu\nu}^a \tilde{F}^{a\,\mu\nu}
 \right\}\,,
\label{redqed}
\end{eqnarray}
where the gauge group is assumed to be U(2), $F_{\mu\nu}^a$ and $F_{\mu\nu}$
are the SU(2) and U(1) gauge field tensors, with the coupling constants
$g_2$ and $g_1$, respectively, $\xi$ is a constant of dimension $m^2$
triggering the Higgsing of the theory, $\theta$ is the vacuum angle, and, finally,
the field $\Phi$ is a 2$\times$2 matrix
 $$\Phi =\{\varphi^{kA}\}$$
where $k$ is the SU($N$) gauge index while $A$ is the flavor
index, $k, A = 1,2$. On the world sheet we get the CP(1) model 
\begin{equation}
{\mathcal L}_{1+1} = \frac{1}{(1+\bar\phi\phi )^2} \left(\, \frac{2}{g^2\, } \partial_\alpha\bar\phi\partial^\alpha \phi
+\frac{\theta}{2\pi\,i}\,\varepsilon^{\alpha\beta}\partial_\alpha\bar\phi\partial_\beta\phi \right )
\end{equation}
plus a free field theory for translational
moduli.
Non-Abelian strings and bulk four-dimensional theories which support them are discussed in detail
in the book \cite{Trev} to which I refer the interested reader. In this talk I will focus
on applications involving axion-like particles.

\section{Axion on the string}
\label{aos}

In the first part I will consider models with ``axion'' localized on the string, and the impact of such axion.

\subsection{The simplest  model}
\label{tsm}

%\begin{figure}
%  \includegraphics[height=.3\textheight]{Pierre}
%  \caption{Picture to fixed height}
%\end{figure}

In the simplest scenario axions (a massless or nearly massless field 
defined on the circle) is the only modulus on the string (except the translational moduli, of course).
The simplest model of this type is obtained from Witten's superconducting string model
\cite{wit1} by its reduction. Namely, we will downgrade one of two U(1)'s of the Witten model 
to a global symmetry, rather 
than local,
\beqn
{\mathcal L} &=&
 -\frac{1}{4g^2}\,F_{\mu\nu}F^{\mu\nu} +|\D_\mu\phi |^2-\frac{\lambda_\phi}{4}\,
\left(\phi^2-v_\phi^2
\right)^2
\nonumber\\[2mm]
&+&
|\p_\mu\chi |^2
-\frac{\lambda_\chi}{4}\,
\left(\chi^2-v_\chi^2
\right)^2
-\beta\phi^2\,\chi^2\,.
\label{3}
\eeqn
This model is a crossbreed between those used in \cite{Sh1,Sh2}. 
If the constants $\lambda_{\phi,\chi}$ and $\beta$ are appropriately chosen, the field $\phi $
condenses 
in the vacuum, Higgsing the gauge U(1) symmetry and, simultaneously, stabilizing the field $\chi$. 
Then in the vacuum $\langle\chi\rangle_{\rm vac}=0$ which implies that the global U(1)
associated with the $\chi$ phase rotations remains unbroken.
The theory (\ref{3}) obviously
 supports a string which is almost the Abrikosov string. There is an important distinction, however.
 In the string core $\phi =0$, and the $\beta\phi^2\,\chi^2$ term stabilizing $\chi$
 is switched off. Having $\chi=0$ inside the string is energetically inexpedient. Thus the string solution has
 $\chi \neq0$ in the core \cite{wit1}. This spontaneously breaks the global U(1) on any given string solution.
As a result, a massless phase field $\in {\rm U}(1)$ -- an axion -- is localized on the string. The world-sheet theory 
becomes
\beq
S =\int dt\,dz \, \left\{  T\left[ (\p_\mu x_0 )^2 + (\p_\mu y_0 )^2\right] + f^2 (\p_\mu\alpha )^2
\right\}
\label{4}
\eeq
where $T$ is the string tension, $f$ is a (dimensionless) axion constant which can be expressed
in terms of the bulk parameters, $t$ is time, $z$ is the coordinate along the string
while $x_0$ and $y_0$ are perpendicular coordinates. They can be combined as $x_\perp =\{x_1,\,x_2\}$,
where $x_\perp $ depends on $t$ and $z$, 
$$x_\perp = x_\perp (t,z)\,.$$
Moreover, $\alpha(t,z)$ is the phase field on the world sheet, 
$\alpha \leftrightarrow \alpha \pm 2\pi \leftrightarrow \alpha \pm 4\pi \,...\,$. In other words, the target space of $\alpha$ is the unit circle.

\begin{figure}
  \includegraphics[height=.15\textheight]{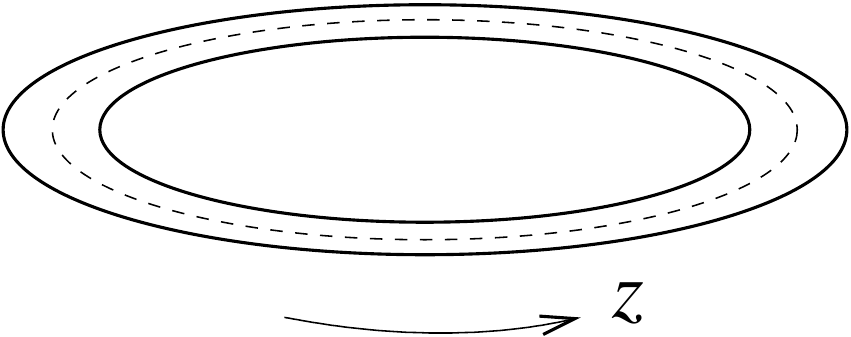}
  \caption{A torus made of the bent Abrikosov flux tube. The dashed line along the
  large period of the torus is the line of constant $\alpha$ (say, $\alpha =0$).}
  \label{prev}
\end{figure}

Now, let us take a long Abrikosov string and and bend it into a circle of circumference $L$, 
see Fig~\ref{prev} (I assume $L\gg \ell$
where $\ell$ is the string thickness). 
If $\alpha$ is constant along $z$ (say, $\alpha = 0$), this configuration is obviously 
unstable. Minimizing its energy, the torus will shrink until $L$ becomes of the order of $\ell$, and then the string will annihilate. However, one can stabilize it by forcing 
$\alpha$ to wind along $z$ in such a way as to make the full $2\pi$ winding when $z$ changes from 0
to $L$,
\beq
\alpha (t,z) = 2\pi z/L\,.
\eeq
Note that $\alpha$ linearly depending on $z$ goes through the equation of motion on the world sheet, 
$\p^2 \alpha =0$.
It is not difficult to estimate the value of $L$. Indeed,
the string energy is
\beq
E = TL + \frac{(2\pi \, f)^2}{L}
\label{6}
\eeq
Minimizing (\ref{6}) with respect to $L$ we get
\beq
L = 2\pi f/\sqrt{T}\,.
\eeq
Making $f$ large enough we can always force $L$ to be much larger than the flux tube thickness
$\ell$ which is roughly speaking of the order of $1/\sqrt{T}$. Note that for $k$ windings
$L = 2\pi f\, k/\sqrt{T}\,$.

The soliton of the type discussed above was first constructed
 in \cite{Sh1} where it goes under a special name ``vorton'' (in the context of cosmic strings;
for a recent review and a rather extended list of references see
\cite{radu}). Its classical stability is due to a nontrivial Hopf topological number \cite{Fadd}.
In the limit $L\gg\ell$ the Hopf soliton is also stable with regards to the
quantum tunneling annihilation.

\begin{figure}
  \includegraphics[height=.22\textheight]{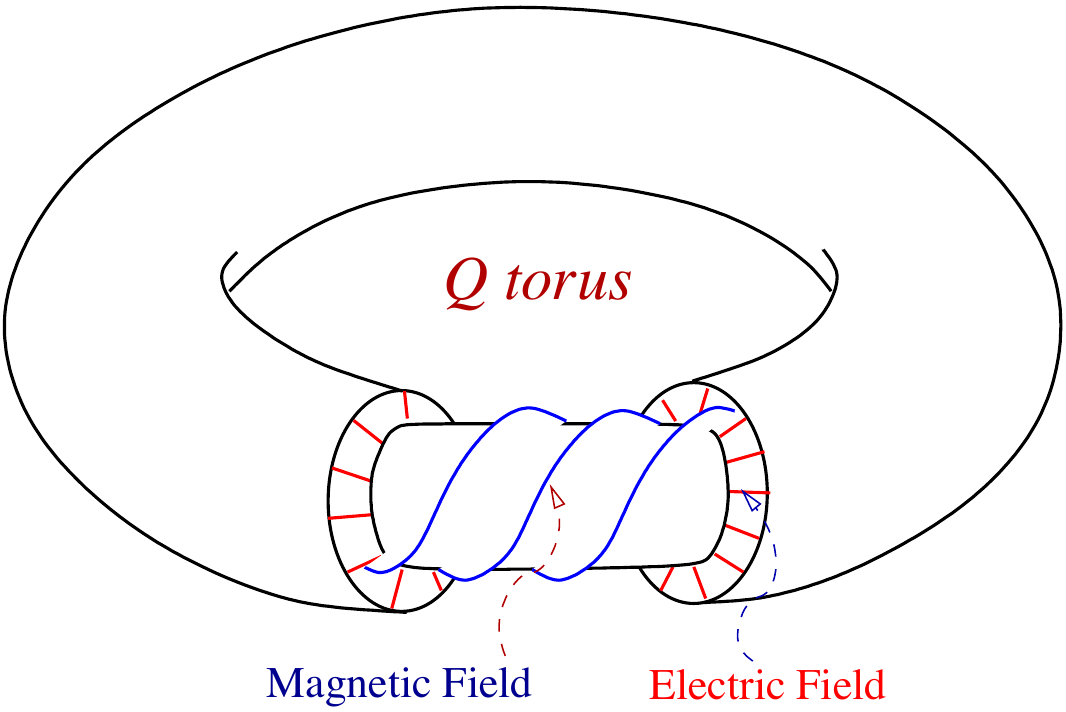}
  \caption{Twisted torus constructed in \cite{BMS}. This Hopf soliton is also topologically stable.}
  \label{potentialp}
\end{figure}

A similar Hopf soliton, albeit with a richer internal structure, was obtained in
\cite{BMS} in the framework of \ntwo supersymmetric QED, see Fig.~\ref{potentialp}.

\subsection{Axion-induced deconfinement of kinks in two dimensions}
\label{aidc}

Now I will pass to the axion impact \cite {Gorsky}
 on ``genuinely'' non-Abelian strings,
with the orientational moduli on the world sheet described by the CP$(N-1)$  model.
In the gauged formulation the CP$(N-1)$ model can be written as
\beq
{\cal L} = \frac{2}{g^2}\, \left[
\left(\partial_{\alpha} + i A_\alpha\right) n^*_{\ell}
\left(\partial_\alpha - i A_\alpha \right) n^{\ell}
-\lambda \left( n^*_{\ell} n^{\ell}-1\right)
\right]\,,
\label{onep}
\eeq
where $n^\ell$ is an $N$-component complex filed, $\ell = 1,2,...,N$, subject to the constraint
 \beq
n_{\ell}^*\, n^{\ell} =1\,.
\label{lambdaco}
\eeq
This constraint is implemented by the Lagrange multiplier $\lambda$
in Eq.~(\ref{onep}). The field $A_\alpha$ in this Lagrangian is also auxiliary,
it enters with no derivatives and can be eliminated
by virtue of the equations of motion,
\beq
A_\alpha =-\frac{i}{2}\, n^*_\ell \stackrel{\leftrightarrow}{\partial_\alpha} n^\ell\,.
\label{two}
\eeq
Substituting Eq.~(\ref{two}) in the Lagrangian, we rewrite it in the form
\beq
{\cal L} = \frac{2}{g^2}\, \left[
\partial_{\alpha}  n^*_{\ell}\,
\partial_\alpha  n^{\ell} + (n_\ell^*\partial_\alpha n^\ell)^2
-\lambda \left( n^*_{\ell} n^{\ell}-1\right)
\right]\,.
\label{three}
\eeq
The coupling constant $g^2$ is asymptotically free, and defines
a dynamical scale $\Lambda$ of the theory  by virtue of the dimensional transmutation,
\beq
\Lambda^2 = M_{\rm uv}^2 \exp\left(-\frac{8\pi}{Ng^2_0}\right)\,,
\label{seven}
\eeq
where $M_{\rm uv}$ is the ultraviolet cut-off and $g^2_0$
is the bare coupling.

At first, let us forget for
a while about the axion  and outline
the solution of the ``axionless''  CP$(N-1)$ model (\ref{onep})
at large $N$ \cite{W79}.
To the leading order it is determined by one loop and can be summarized as follows:
the constraint (\ref{lambdaco}) is dynamically eliminated so that all $N$ fields
$n^\ell$ become independent degrees of freedom with the mass term $\Lambda$.
The photon field $A_\mu$ acquires a kinetic  term
\beq
{\cal L}_{\gamma\,\,\rm kin}= -\frac{1}{4e^2} F_{\mu\nu}^2\,,\qquad e^2 = \frac{12\pi \Lambda^2}{N}\,,
\label{pki}
\eeq
and also becomes ``dynamical.'' The quotation marks here are used
because in two dimen\-sions the kinetic term (\ref{pki})
does not propagate any physical degrees of freedom; its effect
reduces to an instantaneous
Coulomb interaction. This is best seen in the $A_1=0$ gauge.
In this gauge the above kinetic term takes the form $(\partial_z A_0)^2$
%\beq
%(\partial_z A_0)^2
%\eeq
while the interaction is
\beq
A_\alpha J^\alpha =A_0 J^0\,,\qquad J^\alpha=  n^*_\ell \stackrel{\leftrightarrow}{\partial_\alpha} n^\ell\,.
\eeq
Since $A_0$ enters in the Lagrangian without time derivative,
it can be eliminated by virtue of the equation of motion leading
to the instantaneous Coulomb interaction
\beq
J_0 \, \partial_z^{-2}\,  J_0\,.
\label{dev}
\eeq
In two dimensions the Coulomb interaction is proportional to $ |z|\,,$
%\beq
%\frac{\Lambda^2}{N}\,  |z|\,,
%\eeq
implying linear confinement acting between the $n,\,\,\bar n$ ``quarks'' \cite{W79}.
Only $n\,\bar n$ pairs are free to move along the string.

The axion part of the Lagrangian can be written as follows:
\beq
{\cal L}_a = f_a^2\, (\partial_\mu a)^2 +\frac{a}{2\pi }\, \varepsilon_{\alpha\gamma}
\partial^\alpha A^\gamma\,,
\label{eight}
\eeq
where $A^\gamma$ is defined in Eq.~(\ref{two}), and $f_a$ is a (dimensionless)  axion constant.
 I will continue to  assume that $f_a\gg1$.

Bringing kinetic terms to canonical normalization
one obtains
\beq
- \frac{1}{4} F^2_{\mu\nu} + \frac{e}{2\pi f_a}\, a\, \varepsilon_{\alpha\gamma}
\partial^\alpha A^\gamma +
(\pt_\mu a)^2+ e\, A_\alpha J^\alpha\,.
\label{dven}
\eeq
The expression for $e^2$ is given in (\ref{pki}).
The axion field represents a single degree of freedom. The role of the ``photon" is that
upon diagonalization we get a {\em massive} spin-zero particle,
with mass of the order of $f_a^{-1}\Lambda N^{-1/2}$.
Indeed,
taking account of the photon-axion mixing amounts to summing the infinite series of
tree graphs,
\beqn
e^2J^0J^0 \left\{
\frac{1}{p^2} +\frac{1}{p^2}\left(\frac{e}{2\pi f_a}\right)^2\,\frac{1}{p_\mu^2} + ...\right\}=
- e^2J_\alpha J^\alpha\,\, \frac{1}{p_\mu^2 -\left(\frac{e}{2\pi f_a}\right)^2}\,,
\label{mpole}
\eeqn
where $p$ is the spatial component of the momentum transfer $p_\mu$,
and I used  Eqs.~(\ref{dev}) and (\ref{dven}), and the current conservation.
 The ``ex-photon'' mass is determined by the position of the pole in (\ref{mpole}).

As a result,
the long distance force responsible for confinement {\em disappears},
giving place to deconfinement at distances $\gg m_a^{-1}$.

The axion-induced liberation of the $n$ fields at distances $\gg m_a^{-1}$  demonstrated 
above is a two-dimensional counterpart
of domain-wall deconfinement in four-dimensions \cite{Gab2000,GabS}.
The parallel becomes even more pronounced in the (string-inspired)
formalism which ascends to \cite{susskind} (in connection with 
walls it was developed in \cite{Gab2000} and 
discussed in \cite{dvali} in another context). In this  formalism one introduces an
(auxiliary) antisymmetric three-form
gauge field $C_{\alpha\beta\gamma}$, while the
four-dimensional axion is replaced by an
antisymmetric two-form field $B_{\mu\nu}$ (the Kalb--Ramond field).
In four dimensions the gauge three-form field has no propagating
degrees of freedom while the Kalb--Ramond field $B_{\mu\nu}$ presents
a single degree of freedom. 
The domain walls are the sources for $C_{\alpha\beta\gamma}$, much in the same way as the kinks are the sources for $A_0$ in two dimensions. The field strength four-form built from $C_{\alpha\beta\gamma}$
is constant (cf. $F_{01}$ in two dimensions).
The $C_{\alpha\beta\gamma}B_{\mu\nu}$ mixing produces one massive physical degree of freedom, a
four-dimensional massive axion. Simultaneously,
the domain-wall confinement is eliminated at distances  $\gg m_a^{-1}$. Everything is parallel to
 the two-dimensional CP$(N-1)$ world-sheet theory.

\section{Four-dimensional axion and non-Abelian strings}
\label{tri}

Now, I address a  different problem:
a non-Abelian string soliton coupled with a four-dimensional axion existing in the bulk.
We introduce a four-dimensional axion in the bulk theory which supports
non-Abelian strings;  confined monopoles are seen as kinks
in the world-sheet theory (CP$(N-1)$). What's the impact of this four-dimensional axion
on dynamics of strings/confined monopoles?

\subsection{The bulk model with non-Abelian strings}
\label{bmna}

The appropriate bulk theory (nonsupersymmetric)
is given in Eq.~(\ref{redqed})
where $T^a$ stands for the generator of the gauge SU(2) group,
\beq
\nabla_\mu \, \Phi \equiv  \left( \partial_\mu -\frac{i}{\sqrt{ 2N}}\; A_{\mu}
-i A^{a}_{\mu}\, T^a\right)\Phi\, ,
\label{dcde}
\eeq
and $\theta$ is the vacuum angle, to be promoted to the axion field,
\beq
\theta \to \theta +a \to a(x) \,.
\label{promo}
\eeq
 The last term forces $\Phi$ to develop a vacuum
expectation value (VEV) while the last but one term
forces the VEV to be diagonal,
\beq
\Phi_{\rm vac} = \sqrt\xi\,{\rm diag}\, \{1,1\}\,.
\label{diagphi}
\eeq
This VEV results in the spontaneous
breaking of both gauge and flavor SU(2)'s.
A diagonal global SU(2) survives, however,
namely
\beq
{\rm U(2)}_{\rm gauge}\times {\rm SU(2)}_{\rm flavor}
\to {\rm SU(2)}_{\rm diag}\,.
\eeq
The vacuum is color-flavor locked.

One can
combine the $Z_2$ center of SU($2$)  to
get a topologically stable string solution \cite{HT1,ABEKY}
possessing both windings, in SU($2$) and U(1) since $\pi_1 \left({\rm SU}(N)\times {\rm U}(1)/ Z_N
\right)\neq 0\,.$
Their tension is $1/2$ of that of  
the Abrikosov string. It is rather obvious that these strings have
 orientational zero modes
associated with rotation of their color flux inside the non-Abelian
subgroup SU($2$) of the gauge group \cite{Trev}. 
This implies that the effective low-energy theory on the string world sheet
includes both the standard  Nambu--Goto action associated with translational
moduli and a sigma model part which describes internal dynamics of the orientational
moduli,
CP$(N-1)$. The four-dimensional axion interaction is
added in (\ref{onep}) through the term
\beq
-
\frac{\theta +a}{2\pi}\,\,\varepsilon^{nk}\,\partial_n\, n^*\partial_k\, n
\,,
\label{cpN}
\eeq
where $\theta$ coincides with the four-dimensional $\theta$
while $a(t,\vec x )$ is the four-dimensional
pseudoscalar field propagating in the bulk.

\subsection{Monopole-antimonopole ``mesons" vs. axion clouds}
\label{mam}

What happens with the
monopole-antimono\-pole meson on the non-Abelian string in the presence of
the four-dimensional axion?
Given the discussion above one might suspect that the
four-dimensional axion induces deconfinement of  monopoles
localized on the non-Abelian string, much in the same way
as the two-dimensional axion. Now I will argue that this does not happen.

The classical action of the four dimensional bulk axion field is
\beq
L_{a}=\int d^4 x \left[ f_{a}^2(\partial a)^2 + \frac{ia}{32\pi^2}
F^{a}_{\mu\nu}\tilde{F}^{a}_{\mu\nu}\right]\,,
\eeq
where in the case at hand $f_a$ has dimension of mass. The axion  has a
small mass generated by four-dimensional bulk instantons
\beq
m_a^2\sim \frac{\Lambda^4_4}{f_a^2}\,\left(\frac{\Lambda_4}{\sqrt{\xi}}
\right)^{b-4}\,,\qquad f_a\gg \Lambda_4\,,
\label{axionmass}
\eeq
where $b$ is the first coefficient of the $\beta$ function in the
theory (\ref{redqed}), but this mass plays no role in what follows. 

The impact of the  bulk axion on the non-Abelian string is two-fold.
First, the axion gets coupled  to the translational
moduli of the string. Assuming that
the string collective coordinates adiabatically depend on the world-sheet coordinates
we  get for this coupling
\beq
{\cal L}_a^{(1)}\sim \xi\int d^4 x\, a(x)\,
\varepsilon^{ij}\,\varepsilon_{\alpha\beta}\,
\partial_{i}x^{\alpha}_\perp \, \partial_{j}x^{\beta}_\perp \,\,
\delta^{(2)}(x-x_{\rm string}(t,z)),
\label{bterm}
\eeq
where the indices $i,j= 0,3$ run over the string world sheet coordinates
 while the indices
$\alpha,\beta =1,2$ are orthogonal to the string world-sheet. (One could  rewrite this expression
 in a covariant form trading the axion field for the Kalb--Ramon two-index field
$B_{\mu\nu}(x)$ but this is not necessary.) The coupling
(\ref{bterm}) is not specific for non-Abelian strings, it
 is generated in the case of the Abrikosov strings as well.

 Now, let us discus orientational moduli.
It is easy to see that   no mixed $n$-$x_\perp$ terms appear in the axion Lagrangian
(at least, in the  the quadratic order in derivatives).
The bulk axion  generates a quadratic in $n$ coupling, as is clearly
seen from Eq.~(\ref{cpN}).
The impact of this term in the axion Lagrangian can be summarized as follows:
\beq
{\cal L}_a^{(2)}\sim\int d^4x\,a(x)\,
\varepsilon^{nk}\,\,\pt_n\, n^*\pt_k\, n\,\, \delta^{(2)}(x-x_{\rm string}(t,z)),
\label{axionpot}
\eeq

For a short while forget about axions and
consider the monopole-antimonopole pair attached to the string.
The energy of this monopole-antimonopole meson is of order of
$
(\Lambda^2/N)\,L,
$
where $L$ is the distance between the monopole and antimonopole along the string.
What happens upon 
switching on the four-dimensional axion field?

\begin{figure}
  \includegraphics[height=.15\textheight]{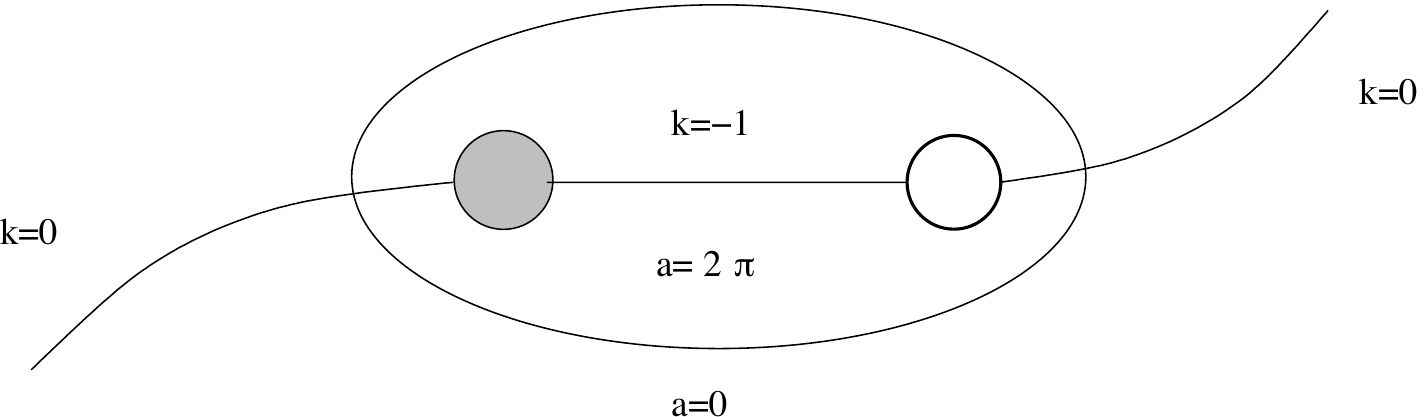}
  \caption{The
monopole-antimonopole meson together with the axion cloud. The region
of the string between the
monopole and antimonopole is not exited because  
the  value of the axion field is nonvanishing inside the axion cloud, $a=2\pi$.}
  \label{fig:cloud}
\end{figure}

Logically speaking, 
the axion field could develop a nonvanishing expectation value
$a=2\pi$ on the string between the monopole and antimonopole positions,
equalizing the string energies inside and outside the pair, and 
screening the confinement force. This is exactly what happened with
the two-dimensional
axion.
To see whether or not a similar effect occurs with four-dimensional
axions we have to examine a field configuration in which $\langle a\rangle =0$
everywhere in the bulk except a region adjacent to the
monopole-antimonopole separation interval, as depicted in Fig.~\ref{fig:cloud}.
We have to check the energy balance assuming there is an axion cloud
such that on the string inside the
monopole-antimonopole separation interval $\langle a\rangle =2\pi$,
which would let (anti)monopoles move freely along the string, with no confinement
along the string.

It is not difficult to estimate the energy of the axion cloud.
The transverse size of the cloud (in two directions perpendicular to
the string) must be of order of $m_a^{-1}$. The longitudinal dimension is $L$, see
Fig.~\ref{fig:cloud}.
Assuming that $L\gg m_a^{-1}$  we get
\beq
E_{\rm cloud}\sim f_a^2 \,L,
\label{axioncloud}
\eeq
to be compared to the energy
$(\Lambda^2/N)\,L$ of the monopole-antimonopole meson.
Since $f_a$ is supposed to be very large compared to $\Lambda$
we see that the energy of the
axion cloud (\ref{axioncloud}) is much larger than the monopole-antimonopole meson energy.
Developing a compensating axion cloud is energetically
disfavored.
Therefore we conclude that there is no monopole
deconfinement driven by four-dimensional axions.

\subsection{Cosmic non-Abelian string and axion emission}
\label{cnas}

Hashimoto and  Tong
 suggested \cite{Hashimoto} to consider non-Abelian strings
as cosmic string candidates. It is worth discussing
possible signatures of such non-Abelian strings
in the context of axion physics. Obviously, both, translational and orientational modes
can be excited in
collisions.
In the latter case one can think of production of energetic monopole-antimonopole pairs
attached to the string and bound in mesons by the
confining potential along the string, as I described above.
 On the part of the string between the monopole and antimonopole
 (the kink and antikink) the state of the string is described by
a quasivacuum with $k=-1$. In this state
 \beq
 \langle \varepsilon^{nk}\,\,\pt_n\, n^*\pt_k\, n\rangle \sim \Lambda^2/N\,.
 \eeq
The topological charge density is localized
in the domain of the excited part of the string, and is approximately constant in this domain.
Therefore, as is clear from Eq.~(\ref{axionpot}), this interval,
whose length $L$ oscillates in accordance with the monopole-antimonopole
motion, will serve as a source term in the equation for the axion field.
 Assume that the energy of the
kink-antikink
pair $E\gg\Lambda$ so that they can be treated quasiclassically.
The distance $L$ between the kink and antikink will oscillate between
$-L_0$ and $L_0$ where $L_0\sim  E/\Lambda^2$ with the frequency
$\omega \sim \Lambda^2/ E$,
\beq
L(t) = L_0 \, e^{i\omega t}\,.
\eeq
Therefore, for a distant observer the monopole-antimonopole
meson is seen as a point-like source with the
interaction term
\beq
{\Lambda^2}\, \int d^4 x \,a(x)\,L(t)\,\delta^{3}(r-r_0),
\label{pointsource}
\eeq
where $r_0$ is a position of the meson on the string. The intensity of the
 axion radiation
from this point-like source can be estimated as
\beq
I_a \sim \omega^2\, \frac{\Lambda^4 L^2_0}{f_a^2}\,\frac1{r^2}
\sim \omega^2\, \frac{E^2}{f_a^2}\,\frac1{r^2} \,,
\label{mesonrad}
\eeq
where $r$ is the distance to the observer.

Of course the string produces axion radiation also due to coupling
 with translational modes, Eq.~(\ref{bterm}). This radiation is seen as coming from
a linear source, and can be estimated (per unit length)  as
\beq
I_a\sim \frac{\xi^{1/2}}{f_a^2}\,\frac{E^2}{\ell^2}\,\frac{1}{\rho}\,.
\label{trrad}
\eeq
Here $\rho$ is the  distance from the string to the observer in the plane
orthogonal to the string,
$E$ is the total excitation energy
and $\ell$ is the length of the excited part of the string. 
This radiation is not specific for non-Abelian
strings. Abelian strings produce this radiation as well.

We see that  the non-Abelian string is seen by a distant observer as a
linear source of the axion radiation (\ref{trrad}), with additional point-like
sources of the axion radiation (\ref{mesonrad}) located on the linear source
at the positions of the monopole-antimonopole
mesons.
The rate of the axion radiation depends of $f_a$. The oscillating
kink-antikink pair will shake off energy until exhaustion. The time duration of the
monopole-antimonopole meson de-excitation 
can be estimated as $T\sim E^2 f_a^2$.

\section{Conclusions}

The existence of axion-like particles is almost unavoidable
in the framework of string theory. The impact of axions on various field-theoretic strings
(flux tubes) is multifaceted. 
Two-dimensional axions can stabilize toric Hopf solitons, 
liberate kinks, confined in the absence of axion, and do other equally remarkable jobs.
Introducing a bulk axion in
the ``benchmark" nonsupersymmetric model \cite{Gorsky2} supporting
non-Abelian strings we observe that
the four-dimensional axion does not lead to monopole deconfinement.
In the context of
cosmic strings, the axion emission due to excitations of non-Abelian strings occurs
in a different way compared to that from Abelian string.
Energetic pairs of confined and oscillating 
(anti)monopoles act as an additional  pointlike source  specific to
non-Abelian strings.

In coclusion, I would like to mention that a new study of toric Hopf solitons
stabilized by axion-like fields is under way now \cite{GSYnew}.

\begin{theacknowledgments}
I am grateful to S. Bolognesi, A. Gorsky and A. Yung
for  useful discussions and collaboration. I would like to thank A. Feldshteyn
for providing Fig.~\ref{loto}.
This work   was supported in part by DOE grant DE-FG02-94ER408. 

\end{theacknowledgments}

\bibliographystyle{aipproc}   % if natbib is available

\bibliography{sample}

\IfFileExists{\jobname.bbl}{}
 {\typeout{}
  \typeout{******************************************}
  \typeout{** Please run "bibtex \jobname" to optain}
  \typeout{** the bibliography and then re-run LaTeX}
  \typeout{** twice to fix the references!}
  \typeout{******************************************}
  \typeout{}
 }

\end{document}